\documentclass[11pt,oneside]{amsart}

\usepackage{amsmath}               % AmSLaTeX
\usepackage{times}
\usepackage{hyperref}

\numberwithin{equation}{section}

\title%
{Dynamics and Thermodynamics of Blackholes and Naked
Singularities}

\author{Lorenzo Fatibene}
\address{Dipartimento di Matematica, Universit\`a di Torino, Italy}
\email{fatibene@dm.unito.it}

\author{Mauro Francaviglia}
\address{Dipartimento di Matematica, Universit\`a di Torino, Italy}
\email{francaviglia@dm.unito.it}

\author{Roberto Giamb\`o}
\address{Dipartimento di Matematica e Informatica,
Universit\`a di Camerino, Italy}
\email{roberto.giambo@unicam.it}
\urladdr{http://dmi.unicam.it/\~{}giambo}

\author{Giulio Magli}
\address{Dipartimento di Matematica, Politecnico di Milano, Italy}
\email{magli@mate.polimi.it}

\theoremstyle{plain}
\theoremstyle{plain}
\theoremstyle{plain}
\theoremstyle{plain}
\theoremstyle{definition}
\theoremstyle{remark}
\theoremstyle{definition}

\theoremstyle{plain}

\begin{document}

%\begin{abstract}

%\end{abstract}

\maketitle

\section{Introduction}\label{sec:intro}

The international Workshop on ``Dynamics and Thermodynamics of
Blackholes and Naked Singularities`` took place at the Department
of Mathematics of  the Politecnico of Milano from 13 to 15 May
2004. Participation was restricted to 70 scientists due to
organizational reasons. The workshop was attended by people coming
from several different Countries. Sponsors of the conference were MIUR, SIGRAV
and INdAM-GNFM. The workshop has been a fruitful occasion of
scientific exchange between people interested in various aspects
of blackhole theory, especially Thermodynamics and Gravitational
Collapse - Cosmic Censorship. The main speakers, in alphabetical
order, were: Hakan Andreasson (G\"oteborg), Fernando de Felice
(Padova), Mauro Francaviglia (Torino), Valeri Frolov (Alberta),
Jos\'e Garcia (Madrid), Fabio Giannoni (Camerino), Pankay Joshi
(Tata, Bombay), Jerzy Kijowski (Warsaw),  Dietmar Klemm (Milano),
Brien Nolan (Dublin), Luciano  Rezzolla (Sissa, Trieste), Louis
Witten (Cincinnati). Two sessions of contributed talks took place
as well. Chairpersons of the session were Lorenzo Fatibene and
Roberto Giamb\`o. Proceedings of the Conference have now been
e-published on the website of the Department
\href{http://www.mate.polimi.it/bh}{http://www.mate.polimi.it/bh}, with a mirror copy
on the SIGRAV website. Each paper is available free of charge
following the links given below.

\bigskip

\section{Contribution list}

\subsection{Invited speakers}

\begin{enumerate}

\item{Gianluca Allemandi, Lorenzo Fatibene, Marco Ferraris,
Mauro Francaviglia, Marco Raiteri, \emph{Geometric Framework for Entropy
in General
Relativity}}\ \ \href{http://www2.mate.polimi.it/convegni/viewabstract.php?id=37&cf=6}{$\longrightarrow$}

\item{H\aa kan Andr\'easson,
\emph{Analytical and numerical results on the
Einstein-Vlasov collapsing system}}\ \
\href{http://www2.mate.polimi.it/convegni/viewabstract.php?id=38&cf=6}{$\longrightarrow$}

\item{Ewa Czuchry,
Jacek Jezierski, Jerzy Kijowski, \emph{Thermodynamics of black holes from
the Hamiltonian point of view}}\ \
\href{http://www2.mate.polimi.it/convegni/viewabstract.php?id=39&cf=6}{$\longrightarrow$}

\item{Fernando de Felice, \emph{Gamma Ray Burst and Cosmic
Time Machines}}\ \
\href{http://www2.mate.polimi.it/convegni/viewabstract.php?id=40&cf=6}{$\longrightarrow$}

\item{Valeri P. Frolov, \emph{Black holes in a spacetime
with large extra dimensions}}\ \
\href{http://www2.mate.polimi.it/convegni/viewabstract.php?id=41&cf=6}{$\longrightarrow$}

\item{Fabio Giannoni, \emph{Non-linear o.d.e. techniques in
gravitational collapse
and Cosmic Censorship}}\ \
\href{http://www2.mate.polimi.it/convegni/viewabstract.php?id=42&cf=6}{$\longrightarrow$}

\item{{Pankaj S. Joshi}, \emph{Gravitational Collapse End
States}}\ \
\href{http://www2.mate.polimi.it/convegni/viewabstract.php?id=43&cf=6}{$\longrightarrow$}

\item{Dietmar Klemm, \emph{Blackholes and singularities in
string theory}}\ \
\href{http://www2.mate.polimi.it/convegni/viewabstract.php?id=44&cf=6}{$\longrightarrow$}

\item{Jos\'e M. Mart\`{\i}n-Garc\`{\i}a, \emph{Critical
Phenomena
and the
global structure of the Choptuik spacetime}}\ \
\href{http://www2.mate.polimi.it/convegni/viewabstract.php?id=45&cf=6}{$\longrightarrow$}

\item{Brien Nolan, \emph{Singularity Strengths and Extendibility}}\ \
\href{http://www2.mate.polimi.it/convegni/viewabstract.php?id=46&cf=6}{$\longrightarrow$}

\item{Louis Witten, \emph{Canonical Quantization of
Collapsing Dust}}\ \
\href{http://www2.mate.polimi.it/convegni/viewabstract.php?id=47&cf=6}{$\longrightarrow$}

\end{enumerate}

\subsection{Contributed papers}

\begin{enumerate}

\item{Emanuele Berti, \emph{A walk in Wonderland: highly
damped black hole quasinormal modes}}\ \
\href{http://www2.mate.polimi.it/convegni/viewabstract.php?id=26&cf=6}{$\longrightarrow$}

\item{Alexander Burinskii, \emph{Singular Strings in the
Rotating Astrophysical Sources}}\ \
\href{http://www2.mate.polimi.it/convegni/viewabstract.php?id=27&cf=6}{$\longrightarrow$}

\item{Marco M. Caldarelli, \emph{On Supersymmetric Solutions
of $D = 4$
Gauged Supergravity}}\ \
\href{http://www2.mate.polimi.it/convegni/viewabstract.php?id=28&cf=6}{$\longrightarrow$}

\item{Roberto Casadio, \emph{Electromagnetic waves on
dilatonic star backgrounds}}\ \
\href{http://www2.mate.polimi.it/convegni/viewabstract.php?id=29&cf=6}{$\longrightarrow$}

\item{Tapas Kumar Das, \emph{Transonic Black Hole Accretion
as Analogue System}}\ \
\href{http://www2.mate.polimi.it/convegni/viewabstract.php?id=30&cf=6}{$\longrightarrow$}

\item{Tomohiro Harada, \emph{Spherically symmetric perfect
fluid collapse in
area-radial coordinates}}\ \
\href{http://www2.mate.polimi.it/convegni/viewabstract.php?id=31&cf=6}{$\longrightarrow$}

\item{Deborah Konkowski, Cassidi Reese, T.M. Helliwell,  C.
Wieland, \emph{Classical and Quantum Singularities of
Levi-Civita Spacetimes with and without a
Positive Cosmological Constant}}\ \
\href{http://www2.mate.polimi.it/convegni/viewabstract.php?id=32&cf=6}{$\longrightarrow$}

\item{Akihiro Ishibashi, \emph{On the stability of black
holes and naked singularities
in static spacetimes}}\ \
\href{http://www2.mate.polimi.it/convegni/viewabstract.php?id=35&cf=6}{$\longrightarrow$}

\item{Hideki Maeda, Takashi Torii, Tomohiro Harada,
\emph{Horizon Instability: a Local Analysis}}\ \
\href{http://www2.mate.polimi.it/convegni/viewabstract.php?id=34&cf=6}{$\longrightarrow$}

\item{{Daniele
Malafarina}, \emph{Physical Properties of the Sources of the Gamma Metric
}}\ \ \href{http://www2.mate.polimi.it/convegni/viewabstract.php?id=33&cf=6}{$\longrightarrow$}

\item{Elizabeth Winstanley, \emph{Classical and
thermodynamical aspects of
black holes with conformally coupled scalar
field hair}}\ \
\href{http://www2.mate.polimi.it/convegni/viewabstract.php?id=36&cf=6}{$\longrightarrow$}

\end{enumerate}

\vspace{1truecm}

\textbf{The whole Proceedings volume is available following
\href{http://www2.mate.polimi.it/convegni/papers/acta_bh.pdf}{this link}
}
\vspace{1truecm}

\end{document}